%% file: penta3loopPRD.tex
\documentclass[aps,prd,superscriptaddress,showpacs,floatfix]{revtex4-2}

\usepackage{graphicx} % Include figure files
\usepackage{dcolumn} % Align table columns on decimal point
\usepackage{bm} % bold math
\usepackage{xspace}
\usepackage{amsmath,amssymb,amsfonts}

\input{commands}
\graphicspath{{Figs/}}

\begin{document}
	
	% Заголовок
	\title{Five legs @ three loops: slightly off-shell dual conformal integrals}
	
	% Авторы и аффилиации
	\author{Andrei V. Belitsky}
	\affiliation{Department of Physics, Arizona State University, Tempe, Arizona 85287-1504, USA}
	
	\author{Leonid V. Bork}
	\affiliation{Dukhov Automatics Research Institute, 127055 Moscow, Russia}
	
	\author{Roman N. Lee}
	\affiliation{Budker Institute of Nuclear Physics, 630090 Novosibirsk, Russia}
	\email{r.n.lee@inp.nsk.su} % Email для корреспонденции
	
	\author{Andrei I. Onishchenko}
	\affiliation{Joint Institute for Nuclear Research, 141980 Dubna, Russia}
	
	\author{Vladimir A. Smirnov}
	\affiliation{Skobeltsyn Institute of Nuclear Physics, Moscow State University, 119992 Moscow, Russia}
	\affiliation{Moscow Center for Fundamental and Applied Mathematics, 119992 Moscow, Russia}
	
	% Аннотация
	\begin{abstract}
		We calculate the three-loop master integrals contributing to the three-loop five-point amplitude on the special Coulomb branch of $\mathcal{N}=4$ SYM theory. For the genuine pentagon integrals, we follow the approach of Ref. \cite{Bork:2025ztu}, which includes a regularization preserving dual conformal invariance (DCI). As a new ingredient, we introduce a simple method, allowing to factor out the dependence on the DCI cross ratios from the contribution of each region. The remaining integrals are then essentially simplified by taking successive limits of vanishing external invariants. For 3 out of 82 regions contributing to the most complicated integral $\mathcal{I}_5^{(3)}$ we were not able to perform the integration even after these simplifications. For these three regions, we perform the integration-by-parts (IBP) reduction in parametric representation and evaluate the resulting locally finite integrals using \texttt{HyperInt}.
	\end{abstract}
	
	% Основной текст
	\maketitle
	
\section{Introduction}

Planar $\mathcal{N}=4$  SYM theory remains the model of choice for theoretical physicists to qualitatively study the perturbative QCD for more than two decades now \cite{Arkani-Hamed:2022rwr}. Yet after so many years of intensive research, this model is still capable of surprising researchers. In particular, it was recently discovered within $\mathcal{N}=4$ SYM that there is likely yet another duality relation between correlation functions of operators with large $R$-charge (``heavy'' operators) and scattering amplitudes on a particular domain of the Coulomb branch of the theory \cite{Caron-Huot:2021usw,Belitsky:2025bgb}. This duality has proved to be extremely important for understanding the IR physics in the vicinity of the mass shell, both in $\mathcal{N}=4$  SYM \cite{Bork:2022vat,Belitsky:2022itf,Belitsky:2023ssv,Belitsky:2024agy,Belitsky:2024dcf,Belitsky:2024yag,Belitsky:2025bez} and in more realistic gauge theories such as QCD \cite{Korchemsky:1988hd}. The key ingredient for investigating this duality is our ability to analytically evaluate dual conformal invariant (DCI) Feynman integrals in a slightly off-shell limit \cite{Caron-Huot:2021usw,Bork:2022vat,Belitsky:2025bgb}. These are planar loop integrals in $d=4$ with massless propagators, which are invariant with respect to conformal transformations of dual variables. The famous example of such integrals is Usyukina-Davydychev ladder integrals \cite{DUfunctions1,DUfunctions2}. The slightly off-shell limit implies that the external legs of the corresponding diagrams have momenta $p_i$ with $p_i^2\neq 0$ being small compared to all other invariants. Apart from these results, there were next to no analytical results available in the literature \cite{Belitsky:2025sin} for such integrals with number of external legs $n>4$, despite lots of efforts. It is important to note that the application of such integrals lies beyond this particular correlator/amplitude duality and can be extended to other slightly off-shell ``observables'' in QCD.

Recently, some of us \cite{Bork:2025ztu} proposed a new approach for evaluating these integrals in a slightly off-shell limit. The key idea of this approach is to retain the DCI symmetry manifest when applying the standard method of regions (MofR) for asymptotic expansion of loop integrals. To this end, a special combination of dimensional and analytic regularizations was used. This combination, on one hand, is sufficient to separate the contribution of different regions, and, on the other hand, preserves the DCI symmetry. This approach significantly simplified the contribution of each region. In particular, it allowed us to calculate the slightly off-shell pentabox integral in terms of logarithms of DCI cross ratios. This integral served as an essential ingredient for the calculation of the two-loop five-point amplitude on the Coulomb branch of planar $\mathcal{N}=4$ SYM \cite{Belitsky:2025bgb}.

\begin{figure}
	\centering
	\begin{tabular}{ccc}
		$\underset{\cI13}{\includegraphics[scale=0.5]{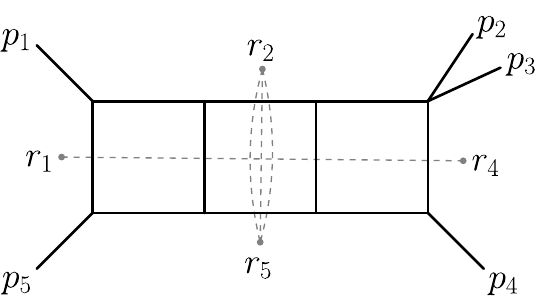}}$
		&
		$\underset{\cI23}{\includegraphics[scale=0.5]{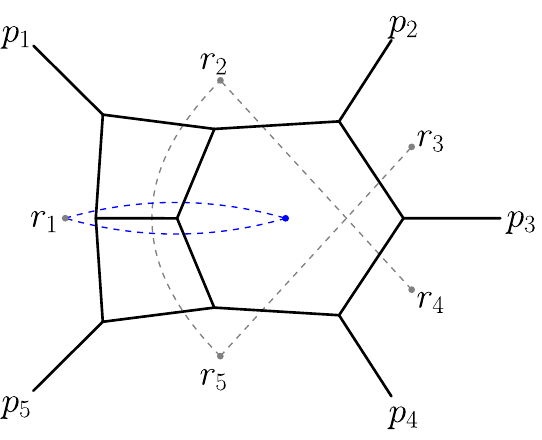}}$
		&
		$\underset{\cI33}{\includegraphics[scale=0.5]{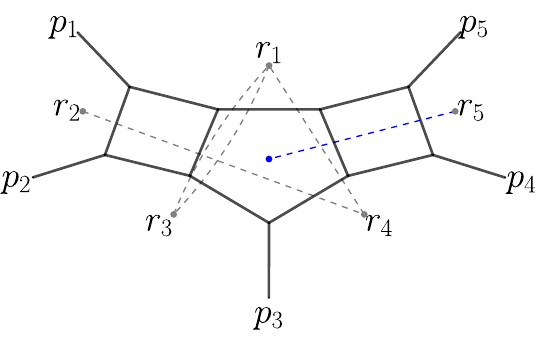}}$
		\\
		$\underset{\cI43}{\includegraphics[scale=0.5]{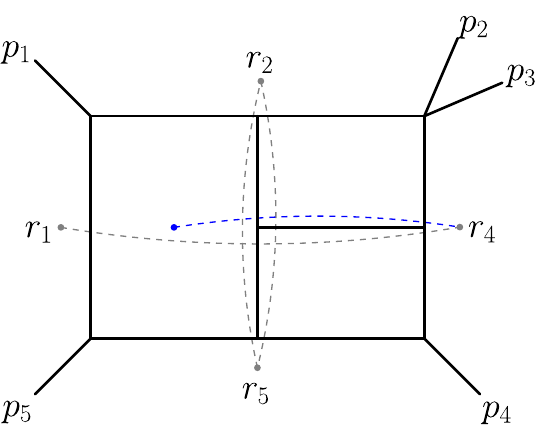}}$
		&
		$\underset{\cI{4'}3}{\includegraphics[scale=0.5]{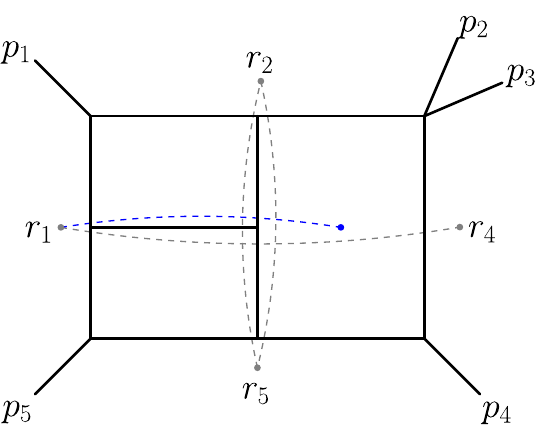}}$
		&
		$\underset{\cI53}{\includegraphics[scale=0.5]{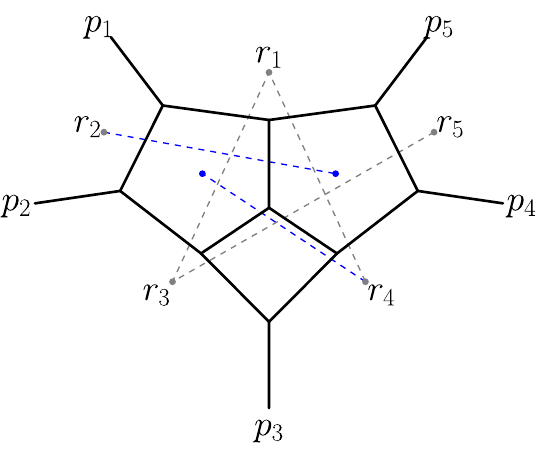}}$
		\\
		$\underset{\cI63}{\includegraphics[scale=0.5]{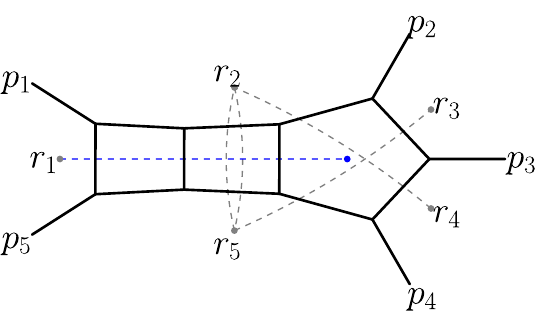}}$
		&
		$\underset{\cI73}{\includegraphics[scale=0.5]{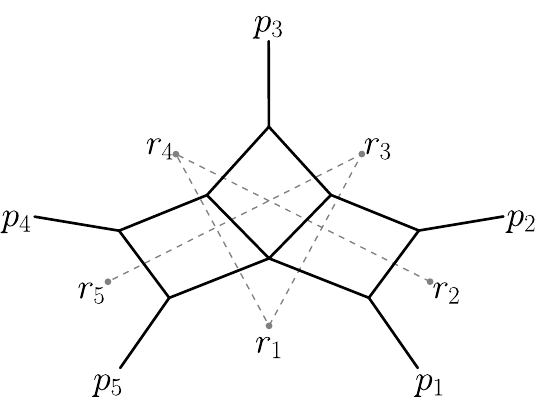}}$
		&
		$\underset{\cI83}{\includegraphics[scale=0.5]{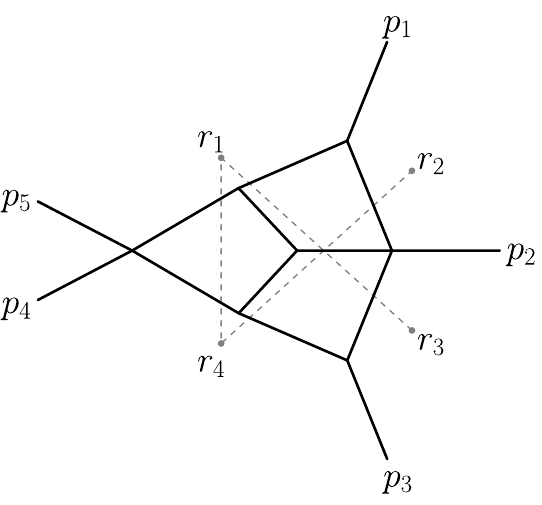}}$
	\end{tabular}
	\caption{Three-loop DCI integrals which appear in the amplitude.}%todo
	\label{fig:DCImasters3}
\end{figure}

In the present paper, we elaborate on the calculation of the three-loop DCI integrals, which contribute to the three-loop five-point amplitude. These integrals are shown in Fig. \ref{fig:DCImasters3}.
We adopt notations from Ref. \cite{penta3l-integrands}, except that we define the integrals without summation over the cyclic and reverse cyclic permutations.

\section{Notations}
Throughout the paper, we use the following notations. The external momenta of the 5-point function are denoted as $p_1,\ldots,p_5$. To avoid confusion with Feynman parameters, we use the notation $r_1,\ldots, r_5$ for dual variables. Then
\begin{equation}\label{kinemdef1}
	p_i =r_{i,i+1},
\end{equation}
where we denote $r_{ij}=r_i-r_j$.
Note that here and below we imply cyclicity in the indices, $i\equiv i+5$.
In general, the pentagon integrals depend on ten kinematic invariants,
\begin{equation}\label{kinemdef2}
	p_i^2 = r_{i,i+1}^2=m_i^2 \text{ and } (p_i+p_{i-1})^2= r_{i-1,i+1}^2=s_i.
\end{equation}
However, these invariants enter the DCI pentagon integrals only via five cross ratios, which we choose as
\begin{equation}\label{DCCvdef}
	v_i =	\frac{r_{i-1,i+1}^2 r_{i-2,i+2}^2}{r_{i-2,i+1}^2 r_{i+2,i-1}^2}
	= \frac{s_i m_{i+2}^2}{s_{i+2}s_{i-2}}
\end{equation}

Note that the equal-mass case $m_i=m$, which is important for the construction of the amplitude, does not reduce the number of independent cross ratios. We however, prefer to hold all masses different for reasons which will be clear below.

When speaking of slightly off-shell integrals, we imply the small-mass asymptotics, when the chosen cross ratios $v_i$ are small compared with unity.

\section{Four point integrals}\label{Boxes}

\begin{figure}
	\centering
	\raisebox{-0.5\totalheight}{\includegraphics[scale=0.4]{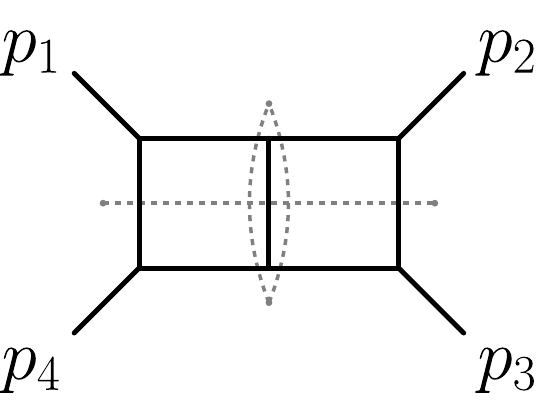}}
	\raisebox{-0.5\totalheight}{$\quad = \quad$}
	\raisebox{-0.5\totalheight}{\includegraphics[scale=0.4]{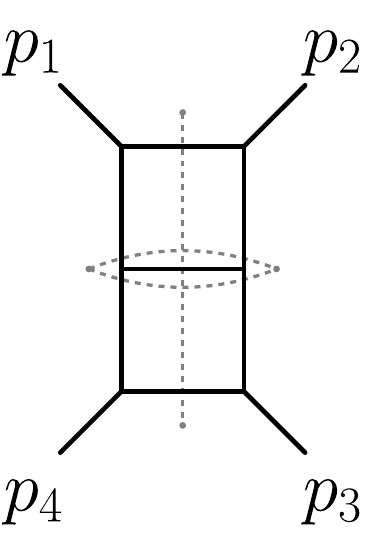}}
	\caption{
		%Magic identity for double boxes.
		Symmetry properties of Usyukina-Davydychev ladder integrals, $L=2$ example.}
	\label{fig:db1=db2}
\end{figure}

Before considering quite complicated five-point three-loop integrals, it is instructive to consider simpler four-point integrals first. In this section, we use notations \eqref{kinemdef1} and \eqref{kinemdef2}.

It turns out that all such integrals, namely $\cI13,\cI43,\cI{4'}3$ and $\cI83$ can be evaluated exactly in $d=4$ for arbitrary values of $s_i$ and $p_i^2$. Let us start our discussion with $\cI13$.

Ladder integrals like $\cI{1}3$ belong to the family of so-called Usyukina-Davydychev ladder integrals pioneered in the famous papers \cite{DUfunctions1,DUfunctions2} by the same authors. The result for the series of these integrals at $L$-loops can be written as:
\begin{equation}\label{BoxFunct}
	\Psi^{(L)}(u,v)=-\sum_{j=L}^{2L}\frac{j!(-1)^{j}}{L!(j-L)!(2L-j)!}
	\log^{2L-j}\left( \frac{v}{u} \right)
	\left(
	\frac{\mbox{Li}_{j}\left(-\frac{1}{\rho~u}\right) - \mbox{Li}_{j}\left( -\rho~v \right)}
	{\lambda}
	\right),
\end{equation}
where
\begin{equation}
	\lambda(u,v)=\sqrt{(1-u-v)^2-4uv},~\rho(u,v)=\frac{2}{1-u-v-\lambda(u,v)},
\end{equation}
and $u,v$ are dual conformal cross ratios defined as:
\begin{equation}\label{4pCCR}
	u = \frac{r_{12}^2 r_{34}^2}{r_{13}^2 r_{24}^2} = \frac{p_1^2p_3^2}{s_{2}s_{3}} ,\quad v =\frac{r_{23}^2 r_{41}^2}{r_{13}^2 r_{24}^2} = \frac{p_2^2p_4^2}{s_{2}s_{3}} \, .
\end{equation}
Sometimes it is more convenient to use a different parametrization of $\Psi^{(L)}$ in terms of $z,\bar{z}$ variables related to $u,v$ as: $u= z\bar{z}$, $v=(1-z)(1-\bar{z})$. This result for the ladder-type integrals is exact in $p_i^2$. Also, it possesses a very important property: it is invariant under $+1$, $mod(4)$ shift of all indices, or in other words, under $s\leftrightarrow t$ Mandelstam variables exchange. The vertical $L$-loop ladder is equal to the horizontal one. See Fig. \ref{fig:db1=db2} for $L=2$ example. Since the corresponding identity holds for arbitrary $p_i^2$, such a transformation can be performed when the $l$-loop ladder integral is a subgraph of another $L$-loop ($l<L$ ) DCI integral. This allows one to generate highly non-trivial relations between DCI integrals known as the \emph{magic identities} in the literature \cite{magic-identities}. At the three-loop level these identities manifest themselves as $\cI{1}3=\cI{4}3=\cI{4'}3$. In words, ``Three loop ladder is equal to tennis court'', where \emph{tennis court} is a nickname for $\cI{4}3$ type integrals \cite{Smirnov:2004ym}.

For $p_i^2=m^2 \ll 1$, $i=1,2,3$ and arbitrary $p_4^2$ we can use the exact result \eqref{BoxFunct} to obtain its small $m$ expansion. Naturally, the results for $\cI{1}3$ and $\cI{1}2$ obtained via the expansion of \eqref{BoxFunct} coincide with those obtained with MofR in DCI regularization. The results of the latter approach for $\cI{1}2$ can be found in \cite{Bork:2025ztu} and those for $\cI{1}3$ will be presented below.

It is instructive to recall how the analytical expression for $\Psi^{(L)}(u,v)$ was originally obtained. Formally, for \emph{any} DCI integral, the result (functional form and the number of arguments) will not change if one considers $r_i \rightarrow \infty$ limit for some particular $r_i$. But on the diagram side, this limit will correspond to the shrinking of one or several propagators. For example, at $L=1$ this will imply that on the functional dependence level, the box with arbitrary $p_i^2$ is equal to the triangle with arbitrary external moments $q_j^2$, $j=1,2,3$. At higher loops, ladder integrals can also be reduced to a triangle ladder. The latter are simpler objects to compute, which allowed Usyukina and Davydychev to uncover their iterative structure and thus to reconstruct four-point $L$-loop ladder integrals \cite{DUfunctions1,DUfunctions2}.
\begin{figure}
	\centering
	\includegraphics[width=0.3\linewidth]{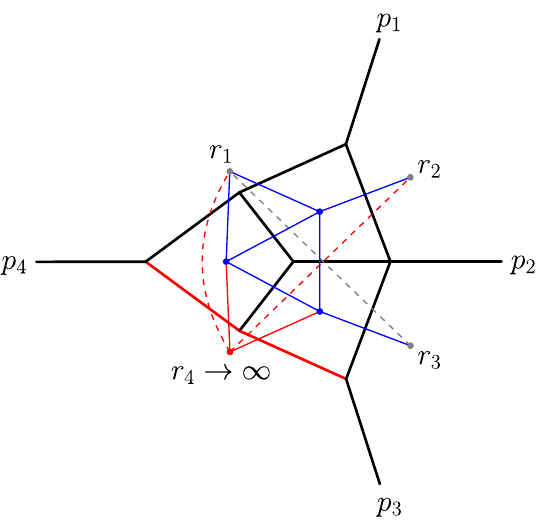}
	\raisebox{2cm}{$\qquad\implies\qquad$}\includegraphics[width=0.3\linewidth]{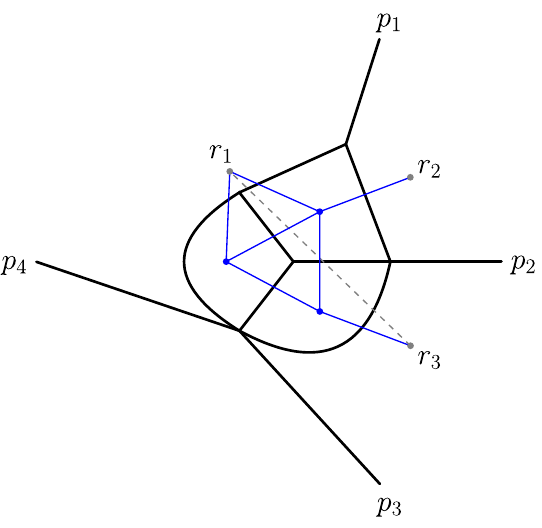}
	\caption{Usyukina-Davydychev trick for $\cI83$ box integral. When we send $r_4$ to infinity, the red lines of the dual graph are removed and those of the conventional graph are contracted, so we obtain the graph on the right.}
	\label{fig:I8int2Triangle}
\end{figure}
The trick with taking one point to infinity is valid for any DCI integral and, in particular, can be applied to $\cI{8}3$. This integral is also a function of the same $u,v$ cross ratios as the Usyukina-Davydychev $L$-loop ladder. This allows one to reduce $\cI{8}3$ to the triangle integral depicted in Fig. \ref{fig:I8int2Triangle}. All triangles with massless propagators and off-shell legs without numerators are expected to be linearly reducible \cite{Panzer:2014gra}. And in our particular case, this allows us to compute the corresponding integral with \HI \cite{Panzer:2014HypIntShort,Panzer:2015idaHypIntLong}. The result can be cast in terms of HPL's and is given by:
\begin{align}
	\cI{8}3
	&=
	\frac{1}{\bar{z}-z}
	\Big[
	-H_{4,1}(z)H_{0}(\bar{z})
	-H_{0,0,0}(z)H_{3}(\bar z)
	-H_{2,2}(z)H_{2}(\bar z)
	+H_{3,2}(z)H_{0}(\bar z)
	+H_{3,0,0}(z)H_{0}(\bar z)
	-H_{0}(z)H_{3,0,0}(\bar z)
	\nonumber\\
	&
	-H_{0,0}(z)H_{2,1,0}(\bar z)
	+H_{0}(z)H_{4,0}(\bar z)
	+4\zeta_3 H_{2,1}(z)
	+4\zeta_3 H_{3}(z)
	-4\zeta_3 H_{3}(\bar z)
	-H_{2}(z)H_{2,1,0}(\bar z)
	+H_{3}(z)H_{0,0,0}(\bar z)
	\nonumber\\
	&
	+H_{2}(z)H_{2,2}(\bar z)
	-4\zeta_3 H_{2,1}(\bar z)
	-H_{0,0}(z)H_{3,0}(\bar z)
	-H_{0}(z)H_{2,1,0,0}(\bar z)
	+H_{2,1,0}(z)H_{2}(\bar z)
	+H_{2,1}(z)H_{2,0}(\bar z)
	\nonumber\\
	&
	+H_{2,1,0,0}(z)H_{0}(\bar z)
	-H_{2,2}(z)H_{0,0}(\bar z)
	+H_{0}(z)H_{2,2,0}(\bar z)
	+H_{3,0}(z)H_{0,0}(\bar z)
	-H_{0,0,0}(z)H_{2,1}(\bar z)
	-H_{0}(z)H_{2,1,2}(\bar z)
	\nonumber\\
	&
	-H_{2,0}(z)H_{2,1}(\bar z)
	-H_{2,2,1}(z)H_{0}(\bar z)
	+H_{0,0}(z)H_{2,2}(\bar z)
	+H_{2,1,2}(z)H_{0}(\bar z)
	+H_{0,0}(z)H_{4}(\bar z)
	-H_{2,2,0}(z)H_{0}(\bar z)
	\nonumber\\
	&
	-H_{4,0}(z)H_{0}(\bar z)
	+H_{2,1}(z)H_{0,0,0}(\bar z)
	-H_{4}(z)H_{0,0}(\bar z)
	+H_{0}(z)H_{4,1}(\bar z)
	+H_{2,1,0}(z)H_{0,0}(\bar z)
	-H_{0}(z)H_{3,2}(\bar z)
	\nonumber\\
	&
	+H_{0}(z)H_{2,2,1}(\bar z)
	-H_{3,0,0,0}(\bar z)
	+H_{3,2,0}(z)
	-H_{4,1,0}(z)
	-H_{2,2,1,0}(z)
	+H_{2,2,0,0}(\bar z)
	-H_{4,0,0}(z)
	-H_{3,2,0}(\bar z)
	\nonumber\\
	&
	-H_{2,2,0,0}(z)
	+H_{4,0,0}(\bar z)
	+H_{4,1,0}(\bar z)
	+H_{3,0,0,0}(z)
	+H_{2,1,2,0}(z)
	-H_{2}(z)H_{3,0}(\bar z)
	+H_{3}(z)H_{2,0}(\bar z)
	+H_{2}(z)H_{4}(\bar z)
	\nonumber\\
	&
	+H_{3,0}(z)H_{2}(\bar z)
	-H_{2,0}(z)H_{3}(\bar z)
	-H_{4}(z)H_{2}(\bar z)
	-H_{2,1,0,0,0}(\bar z)
	+H_{2,2,1,0}(\bar z)
	-H_{2,1,2,0}(\bar z)
	+H_{2,1,0,0,0}(z)
	\Big] .
\end{align}
Here $z,\bar z$ are related to $u,v$ (\ref{4pCCR}) as $u=z\bar z$ , $v=(1-z)(1-\bar z)$.
The limit $p_i^2=m^2 \ll p_4^2$ ($i=1,2,3$) corresponds to $z \rightarrow 0$, $\bar z \rightarrow 1$. Expanding HPLs, we obtain
\begin{eqnarray}\label{I8ver1}
	\cI{8}3
	&=&
	-\frac{1}{3}\zeta_3 \log^3\!\left(\frac{m^4}{s_{2}s_{3}}\right)
	+3\zeta_4 \log^2\!\left(\frac{m^4}{s_{2}s_{3}}\right)
	-\left(2\zeta_2\zeta_3+10\zeta_5\right)
	\log\!\left(\frac{m^4}{s_{2}s_{3}}\right)
	+\frac{41}{2}\zeta_6-4\zeta_3^2 \nonumber\\
	&+&O(m^2)
	\, .
\end{eqnarray}
In the symmetrical point limit ($s_i=Q^2$ for all $i$), all the coefficients in front of logarithms of  $m$ in \eqref{I8ver1} agree with those in Ref. \cite{Nguyen:2007ya}.
This result can be equivalently rewritten as
\begin{equation}
	\cI{8}3 =
	- \frac{1}{3} (L_1 + L_4)^3 \zeta_3 + 3(L_1 + L_4)^2 \zeta_4 - 2(L_1 + L_4) \zeta_2 \zeta_3 - 10(L_1 + L_4) \zeta_5 + \frac{41}{2} \zeta_6- 4\zeta_3^2 +O(v_i)
	\, ,
\end{equation}
where $L_i=\log v_i$, and $v_i$ is defined according to (\ref{DCCvdef}) if we consider $p_4^2 \mapsto (p_4+p_5)^2=s_5$, $p_i^2=m^2 \ll1$, $i=1,\ldots,5$, which is the relevant kinematics for the amplitude with five external momenta. This result agrees with that obtained below by MofR with the DCI regularization. This can be considered a nice cross-check of our DCI MofR results. This concludes the computation of all necessary four-point integrals contributing to the planar five-point amplitude at $L=3$.

\section{Method of regions in DCI regularization}

\subsection{DCI regularization in $d$ dimensions}\label{sec:DCIreg}

Let us remind the idea of DCI regularization introduced in Ref \cite{Bork:2025ztu}.
Consider the $P$-point $L$-loop DCI integral in $4$ dimensions\footnote{All through the paper we use Euclidean metrics.}
\begin{equation}
	I_{L}(r_1,\ldots r_P)=
	\int \prod_{l=P+1}^{P+L}\frac{d^4r_{l}}{\pi^2}
	\prod_{i=1}^{M} \left(r_{k_i m_i}^2\right)^{-n_i}.
\end{equation}
where $n_i$ are integers and $r_i$ are the standard dual variables. As usual, we use the notation $r_{km}=r_k-r_m$.
The Poincar\'e invariance of this integral is obvious, but the dual conformal invariance also requires that the integral is invariant with respect to the inversion $r_n\to r_n/r_n^2$. In order to derive the precise condition on the indices, let us introduce the indicator function
\begin{equation}
	\theta_{li}=
	\begin{cases}
		1 & \text{if }l\in\{k_i,m_i\}\\
		0 & \text{otherwise}
	\end{cases}
\end{equation}
Then, under inversion, the integrand acquires the factor
\begin{equation}
	\prod_{l=1}^{P} \left(r_l^2\right)^{\sum_in_i\theta_{li}}
	\prod_{l=P+1}^{P+L} \left(r_l^2\right)^{-4+\sum_in_i\theta_{li}}
\end{equation}
Thus, the integral is DCI invariant iff
\begin{equation}\label{eq:dci_constraint_4}
	\sum_in_i\theta_{li}=
	\begin{cases}
		0,&l\leqslant P\\
		4,& l>P
	\end{cases}
\end{equation}

Let us now regularize the integral, both dimensionally and analytically, such that it remains DCI invariant.
In addition to the dimensional regularization parameter $\e=2-d/2$, we introduce the parameters $\a_1,\ldots, \a_M$, so that the regularized integral reads
\begin{equation}\label{eq:Ireg}
	I_{L}^{\text{reg}}(r_1,\ldots r_P)=
	\int \prod_{l=1}^{L}\frac{d^dr_{P+l}}{\pi^{d/2}}
	\prod_{i=1}^{M} \left(r_{k_i m_i}^2\right)^{-\nu_i},
\end{equation}
where $\nu_i=n_i+\a_i$. Let us now discuss restrictions on the choice of $\a_i$. A technical requirement is that we do not want to introduce analytic regularization for numerators that depend on the integration variables. In other words, we put $\a_i=0$ if $\max(k_i,m_i)>P$ and $n_i\leqslant 0$. Next, we have to request the invariance with respect to inversion. This restriction reads
\begin{equation}\label{eq:dci_constraint_d}
	\sum_i\nu_i\theta_{li}=
	\begin{cases}
		0,&l\leqslant P\\
		d,& l>P
	\end{cases}
\end{equation}
Subtracting \eqref{eq:dci_constraint_4} from \eqref{eq:dci_constraint_d}, we obtain the conditions for the regularization parameters
\begin{equation}\label{eq:dci_constraint_d1}
	\sum_i\a_i\theta_{li}=
	\begin{cases}
		0,&l\leqslant P\\
		-2\e,& l>P
	\end{cases}
\end{equation}
In particular, the above equations for $l>P$ mean that one cannot avoid analytic regularization for internal lines, since otherwise the sum in the left-hand side would be zero.

\subsection{Pulling out cross ratios}

Although we know that the DCI preserving regularization secures the DCI invariance of each region, we can not use it as effectively as we did for the hard region in Ref. \cite{Bork:2025ztu}. Therein, we used the fact that the hard region corresponds to an on-shell DCI regularized integral, which should not depend on $s_i$. Then we used successive limits in $s_i\to 0$ to simplify the integrand. What prevents us from the same approach for the contributions of other regions is that they do depend on $s_i$ via the cross ratios. So, if we could pull out of the integral the dependence on the cross ratios, we would be able to apply the same approach to other regions. Let us explain how we do that.

This is exactly why we retained different masses. We write $p_i^2=m_i^2= (m \mu_i)^2$, where $m$ remains a small parameter and $\mu_i=m_i/m$ remains fixed. Then we use MofR which, of course, gives exactly the same regions as for the case of equal masses, in terms of $m$-scaling of Feynman parameters. But now the polynomial in Lee-Pomeransky representation depends on $\mu_i$, and we search for the rescaling $x_k\to \prod_i \mu_i^{\nu_{ik}} x_k$ to make this polynomial homogeneous function of each $\mu_i$. This trick allows us to pull the $m_i$ dependence out of the integral. Then we rewrite $m_i^2$ via $v_{i-2}$ as $m_i^2=\frac{v_{i-2} s_{i} s_{i+1}}{s_{i-2}}$ so as to make the dependence on $v_i$ explicit. At this stage the contribution of each region has the form
\begin{equation}
	\bigg(\prod_kv_k^{\mu_k}\bigg) F(s_1,\ldots, s_5),
\end{equation}
where $\mu_k$ depend on the region and $F(s_1,\ldots, s_5)$ is expressed via parametric integral with nontrivial dependence on $s_i$ in the integrand. However, thanks to DCI we know that $F$ will not depend on $s_i$ after the integration. Then we can apply the same approach as we used for hard region in Ref. \cite{Bork:2025ztu}: we calculate naive successive $s_i\to\infty$ limits of the integrand in order to simplify it.

\begin{figure}
	\centering
	\includegraphics[width=0.7\linewidth]{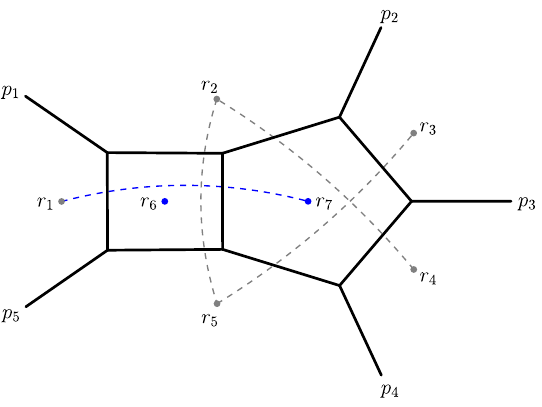}
	\caption{DCI pentabox integral. The dashed lines denote numerators on the dual graph.}
	\label{fig:pentabox}
\end{figure}

Let us explain our approach with a specific example. To make the example simpler and yet informative, we reconsider the calculation of region \#38 of the two-loop slightly off-shell pentagon master integral presented in \cite{Bork:2025ztu}.

The two-loop dual conformal pentabox integral is shown in Fig. \ref{fig:pentabox}. Its DCI regularization reads
\begin{multline}
	PB^{\text{dci}}=PB^{\text{dci}}(\e,\a_1,\ldots, \a_6)\\
	=\int \frac{d^dr_6}{\pi^{d/2}}\frac{d^dr_7}{\pi^{d/2}}
	\frac{r_{14}^{2 \left(-\a_{2\bar 346}\right)} r_{25}^{2 \left(1+\a_{45}\right)} r_{31}^{2 \left(\a_{1\bar 3246}\right)} r_{42}^{2 \left(1+\a_{2\bar 5}\right)} r_{53}^{2 \left(1-\a_{1\bar 324}\right)}r_{17}^2}{r_{56}^{2 \left(1-\a_{12\bar 3}\right)} r_{57}^{2 \left(1+\a_5\right)} r_{61}^{2 \left(1+\a_1\right)} r_{62}^{2 \left(1+\a_2\right)} r_{72}^{2 \left(1+\a_4\right)} r_{73}^{2 \left(1+\a_6\right)} r_{74}^{2 \left(1-\a_{\bar 3456}\right)} r_{67}^{2 \left(1-\a_3-2\e \right)}}
\end{multline}
where we used notations $\a_{ij\ldots k}=\a_i+\a_j+\ldots+ \a_k$, $\a_{\bar n}=-\a_n$. The original integral is recovered as
\begin{equation}
	PB=\left.PB^{\text{dci}}\right|_{\alpha_i,\e\to0}.
\end{equation}

The Lee-Pomeransky parametric representation \cite{Lee:2013hzt,Lee:2014tja} for the regularized integral reads

\begin{equation}
	\label{eq:pentabox_lp}
	PB^{\text{dci}} =
	\intop_{\mathbb{R}_+^8} dx_1\ldots dx_8 f(x) \frac{P(x)}{G(x)^{3-\e}}
\end{equation}
where
\begin{multline}
	G(x) = x_{123} x_{45678} +x_4 x_{5678}
	+s_1
	\left(x_{12} \hat{x}_{56}+\hat{x}_{23} x_{45678}+x_{25}
	\hat{x}_{46}+\hat{x}_{35} x_{46}\right) +s_2
	\hat{x}_{147} \\ +s_3 \left(x_{13}
	\hat{x}_{58}+\hat{x}_{28} x_{45}+\hat{x}_{458}\right) +s_4 \left(x_{12}
	\hat{x}_{67}+\hat{x}_{37} x_{46}+\hat{x}_{46}
	x_7\right) +s_5 \hat{x}_{148}
	+m_1^2 \left(\hat{x}_{12} x_{45678}+\hat{x}_{145}\right)\\
	+m_2^2 \left(x_{1234}
	\hat{x}_{57}+\hat{x}_{247}\right) +x_{1234} \hat{x}_{78}
	m_3^2 +m_4^2 \left(x_{1234}
	\hat{x}_{68}+\hat{x}_{348}\right)+m_5^2 \left(\hat{x}_{13}
	x_{45678}+\hat{x}_{146}\right)\, ,
\end{multline}
\begin{equation}
	f(x) = \frac{\Gamma (3-\epsilon ) x_1^{\alpha _1} x_2^{\alpha _2}x_3^{-\alpha _{12\bar{3}}}
		x_4^{\alpha_{\bar 3}-2 \epsilon } x_5^{\alpha _4} x_6^{\alpha _5} x_7^{\alpha _6}
		x_8^{-\alpha _{\bar{3}456}}
		s_1^{1+\alpha _{45}}
		s_2^{\alpha _{12\bar{3}46}}
		s_3^{1+\alpha _{2\bar{5}}}
		s_4^{1-\alpha_{12\bar{3}4}}
		s_5^{-\alpha _{2\bar{3}46}}
	}{\Gamma \left(1\!+\!\a_1,1\!+\!\a_2,1\!-\!\a_{12\bar 3},1\!+\!\a_4,1\!+\!\a_5,1\!-\!\a_{\bar 3456},1\!+\!\a_6,1\!+\!\a_{\bar 3}\!-\!2\e,\a_{\bar 3}\!-\!1\!-\!\epsilon\right)}\, ,
\end{equation}
and
\begin{multline}
	P(x) = x_{1234} + s_1 \hat{x}_{23} +s_2
	\left(\hat{x}_{17}+x_{234} x_7\right) +s_5
	\left(\hat{x}_{18}+x_{234} x_8\right) \\
	+	m_1^2 \left(x_1 x_{25}+x_2 x_{45}+x_{34} x_5\right)
	+m_5^2 \left(x_1 x_{36}+x_{24} x_6+x_3 x_{46}\right)
\end{multline}
Here we used notations $x_{ij\ldots k}=x_i+x_j+\ldots+ x_k$, $\hat{x}_{ij\ldots k}=x_i\cdot x_j\cdot\ldots\cdot x_k$  and $\Gamma(a_1,\ldots, a_k)=\Gamma(a_1)\cdot\ldots \cdot \Gamma(a_k)$.
Region \#38 corresponds to the rescaling
\begin{equation}
	x_{1,2}\to m^{-2}x_{1,2}
\end{equation}
Retaining only the leading in $m$ term, we get for the contribution of region \#38
\begin{equation}
	PB^{\text{dci}}_{38} =
	\intop_{\mathbb{R}_+^8} dx_1\ldots dx_8 f(x) \frac{P_{38}(x)}{G_{38}(x)^{3-\e}}\, ,
\end{equation}
where
\begin{multline}
	G_{38}(x) = x_1 x_{45678}+x_2
	x_{5678} +\hat{x}_{24} +s_1
	\left(x_6
	\left(\hat{x}_{15}+\hat{x}_{24}+\hat{x}_{25}\right)+\hat{x
	}_{23} x_{45678}\right) +s_2 \hat{x}_{147} \\ +s_3 x_8
	\left(\hat{x}_{15}+\hat{x}_{24}+\hat{x}_{25}\right) +s_4 x_{12}
	\hat{x}_{67}+s_5 \hat{x}_{148} +\hat{x}_{12} x_{45678} \mu_1^2 \label{eq:G1}
\end{multline}
and
\begin{equation}
	P_{38}(x)= s_2 x_{12} x_7+s_5 x_{12} x_8+x_{12}+s_1 \hat{x}_{23} +\hat{x}_{12} \mu_1^2\, .
\end{equation}
We remind that $\mu_i=m_i/m$. In Ref. \cite{Bork:2025ztu}, we proceeded with the explicit variable change and integration to express the contribution in terms of $\Gamma$-functions. But for the three-loop pentagon integrals, which is the main goal of the present paper, the analogue of the function $G_{38}$ appears to be still quite complicated to discover the appropriate change of variables. Therefore, we will show now how we can simplify this form yet further using the fact that each separate contribution is dual conformal invariant.

First, we look for a rescaling of integration variables $x_i\to \prod_{j=1}^5 \mu_j^{a_{ji}}x_i$, such that the dependence on $\mu_j$ in $G_{38}$ polynomial factorizes.
Under this rescaling each monomial of $G_{38}$ transforms as
\begin{equation}
	\prod_{i=1}^8 x_i^{p_i}\prod_{j=1}^5 \mu_j^{n_j}\to \prod_{i=1}^8 x_i^{p_i} \prod_{j=1}^5 \mu_j^{n_j+\sum_{i=1}^8 a_{ji} p_i}.
\end{equation}
Then we require that the factor $\prod_j \mu_j^{n_j+\sum_{i=1}^8 a_{ji} p_i}$ is the same for all monomials. This gives us a system of linear equations for $a_{ij}$. Solving this system, we find the unique solution which corresponds to rescaling $x_{1,2}\to x_{1,2}/\mu_1^2$. Applying this rescaling, we obtain
\begin{equation}
	PB^{\text{dci}}_{38} =m_1^{-\alpha _{12}-\e}
	\intop_{\mathbb{R}_+^8} dx_1\ldots dx_8\, f(x)\left.\frac{P_{38}(x)}{G_{38}(x)^{3-\e}}\right|_{\mu_i=1}.
\end{equation}
Note that at this stage the dependence on the formal small parameter $m$ is reabsorbed into $m_i=m \mu_i$ as expected. Taking into account that $m_1^2=v_4s_1s_2/s_4$, we see that
\begin{equation}
	F=v_4^{\alpha _{12}+\e} PB^{\text{dci}}_{38}=
	(s_1s_2/s_4)^{-\alpha _{12}-\e}	\intop_{\mathbb{R}_+^8} dx_1\ldots dx_8\, f(x)\left.\frac{P_{38}(x)}{G_{38}(x)^{3-\e}}\right|_{\mu_i=1}
\end{equation}
can not depend on cross ratios and is, therefore, constant. Already at this stage, we can conclude that the contribution of this region may provide only logarithms of cross ratios (in fact, of $v_4$ only) when expanded in the regularization parameters. Moreover, we can use consecutive limits $s_i\to 0$ in order to simplify the integrand.
In particular, we use
\begin{equation}
	F=\lim_{s_5\to 0}\lim_{s_4\to 0}\lim_{s_3\to 0}\lim_{s_2\to 0}\lim_{s_1\to 0}F.
\end{equation}
In the asymptotics $s_1\to 0$ we find two different regions, but only one is proportional to $s_1^0$. The corresponding rescaling is given by
\begin{equation}
	x_{3,4,5,6}\to s_1^{-1} x_{3,4,5,6}\,.
\end{equation}
Next, in the limit $s_2\to 0$ we again have two distinct regions. The one contributing to $s_2^0$ is given by rescaling
\begin{equation}
	x_{4,5,7}\to s_2^{-1} x_{4,5,7}\,.
\end{equation}
The limit $s_3\to 0$ has three regions, with again only one contributing to $s_2^0$. The corresponding rescaling is
\begin{equation}
	x_{2,4,5,7,8}\to s_3^{-1} x_{2,4,5,7,8}\,.
\end{equation}
The two remaining limits are done similarly, where for each asymptotics we have only one region contributing to the ``naive'' limit. The total rescaling reads
\begin{equation}
	x_{2}\to \frac{s_5}{s_3} x_2\, \quad
	x_{3,6}\to \frac{x_{3,6}}{s_1} \, \quad
	x_{4,5}\to \frac{s_4 s_5}{s_1 s_2 s_3} x_{4,5}\, \quad
	x_{7}\to \frac{s_5}{s_2 s_3} x_7\, \quad
	x_{8}\to \frac{x_8}{s_3}\, .
\end{equation}
As a result we obtain a much simpler expression for the contribution:
\begin{multline}
	PB_{38}^{\text{dci}}=v_4^{-\alpha_{12}-\epsilon }\Gamma (3-\epsilon )
	\intop_{\mathbb{R}_+^8} dx_1\ldots dx_8  x_1^{\alpha _1} x_2^{1+\alpha _2} x_5^{\alpha _4} x_6^{\alpha
		_5} x_7^{\alpha _6} x_4^{-\alpha _3-2 \epsilon } x_3^{-\alpha
		_{12\bar{3}}} x_8^{-\alpha _{\bar{3}456}}
	\\
	\times\frac{x_{78}\Big(x_{1368} \hat{x}_{24}+\hat{x}_{14} x_{78}+x_2
		\left(\hat{x}_{15}+\hat{x}_{35}+x_{45}+\hat{x}_{67}\right)
		+\hat{x}_{25} x_{68}\Big)^{-3+\epsilon}}{\Gamma \left(1+\a_1,1+\a_2,1-\a_{12\bar 3},1+\a_4,1+\a_5,1-\a_{\bar 3456},1+\a_6,1+\a_{\bar 3}-2\e,\a_{\bar 3}-1-\epsilon\right)}
\end{multline}
The integral in this representation can be easily taken in terms of $\Gamma$-functions by performing rescaling $x_{5,7,8}\to x_4 x_{5,7,8}$ and performing integrations in the sequence $x_1,x_2,x_3,x_6,x_4,x_8,x_7,x_5$.
We obtain
\begin{equation}
	PB_{38}^{\text{dci}}=-
	\frac{\left(\alpha _2+\epsilon \right) \Gamma \left(1+\epsilon +\alpha _{3\bar4},1+\alpha _{2\bar5},-\alpha _{2\bar346},\alpha _{2\bar3}-\epsilon ,\alpha _{\bar345}-\epsilon ,\alpha _{\bar346}-\epsilon ,\alpha _{12}+\epsilon \right)}{v_4^{\alpha_{12}+\epsilon }\Gamma \left(1-\epsilon +\alpha _{2\bar34},1+\alpha _1,1+\alpha _2,1+\alpha _5,1-\alpha _{\bar3456},1+\alpha _6,1-2 \epsilon -\alpha _3\right)}\,,
\end{equation}
which coincides with that presented in \cite[Eq. (4.9)]{Bork:2025ztu}.

\section{Five-point slightly off-shell integrals}

Let us explain in some detail the calculation of $\cI63$.

The DCI regularized version of this integral reads
\begin{multline}
	\cI6{3,\text{dci}}= \int\frac{d^dr_6}{\pi^{d/2}}\frac{d^dr_7}{\pi^{d/2}}\frac{d^dr_8}{\pi^{d/2}}\\
	\times\frac{r_{18}^2 r_{13}^{2 \alpha _{\bar 0\bar 23\bar 6\bar 8}} r_{14}^{2 \alpha _{\bar 0\bar 14\bar 5\bar 7}} r_{24}^{2 \left(\alpha _{01\bar 4578}+1\right)} r_{25}^{2 \left(\alpha _{\bar 034\bar 7\bar 8}+2\right)} r_{35}^{2 \left(\alpha _{02\bar 3678}+1\right)}r_{78}^{2 \left(\alpha _{005678}-1\right)}}{r_{26}^{2 \left(\alpha _1+1\right)} r_{27}^{2 \left(\alpha _3+1\right)} r_{28}^{2 \left(\alpha _5+1\right)} r_{38}^{2 \left(\alpha _7+1\right)} r_{48}^{2 \left(\alpha _8+1\right)} r_{56}^{2 \left(\alpha _2+1\right)} r_{57}^{2 \left(\alpha _4+1\right)} r_{58}^{2 \left(\alpha _6+1\right)}  r_{16}^{2 \left(-\alpha _{0012\bar 3\bar 45678}+1\right)} r_{67}^{2 \left(\alpha _{\bar 3\bar 45678}+1\right)}}
\end{multline}
where we have denoted $\a_0=\e$.

We find 109 regions, and our approach shows that only 54 regions are nonzero.

\begin{figure}
	\centering
	\begin{picture}(300,160)
		\put(0,0){\includegraphics[width=300pt]{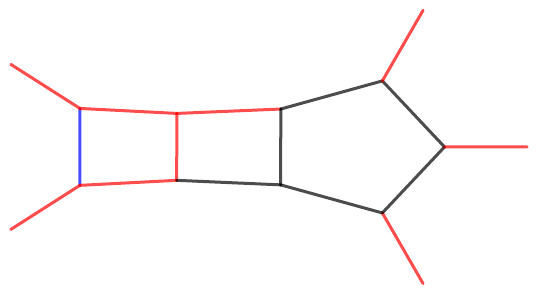}}
		\put(67,106){$1$}
		\put(67,54){$2$}
		\put(124,106){$3$}
		\put(124,54){$4$}
		\put(180,116){$5$}
		\put(180,44){$6$}
		\put(230,103){$7$}
		\put(230,57){$8$}
		\put(37,80){$9$}
		\put(100,80){$10$}
		\put(158,80){$11$}
	\end{picture}
	\label{fig:i6_3_80}
	\caption{One of the regions with a nonzero contribution. Color encodes the virtuality: black, red, and blue color correspond to $m^0$, $m^2$ and $m^4$ virtuality, respectively. Numbers correspond to the indices of Feynman parameters.}
\end{figure}
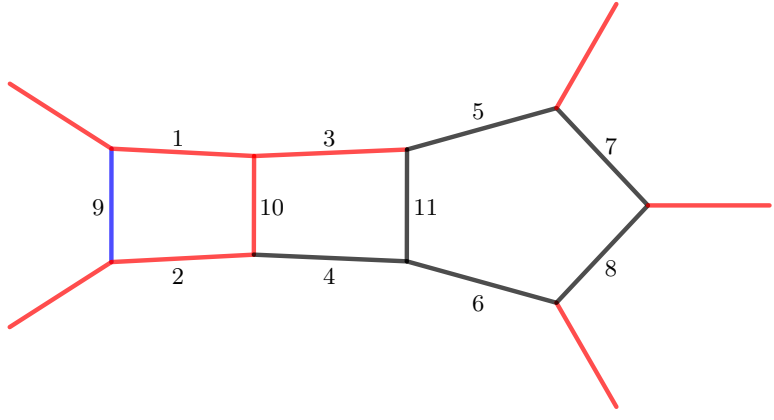

Let us consider, for example, region \#80 depicted in Fig. \ref{fig:i6_3_80}.
Rescaling Feynman parameters and retaining only the leading contribution, we obtain
\begin{equation}\label{eq:I63_80}
	\left[\cI6{3,\text{dci}}\right]_{\#80}=m^{2\alpha _{012\bar4\bar3\bar35678}}
	s_1^{\alpha _{\bar 034\bar 7\bar 8}+2} s_2^{\alpha _{\bar 0\bar 23\bar 6\bar 8}} s_3^{\alpha _{01\bar 4578}+1} s_4^{\alpha _{02\bar 3678}+1} s_5^{\alpha _{\bar 0\bar 14\bar 5\bar 7}}
	\intop_{\mathbb{R}^{11}}dx_1\ldots dx_{9} dx_{A} dx_{B}
	f(x)\frac{P_1(x)}{G_1(x)^{3-\e}}
\end{equation}
where
\begin{equation}
	f(x)= \frac{\Gamma \left(3-\e\right) x_1^{\alpha _1} x_2^{\alpha _2} x_3^{\alpha _3} x_4^{\alpha _4} x_5^{\alpha _5} x_6^{\alpha _6} x_7^{\alpha _7} x_8^{\alpha _8}  x_9^{-\alpha _{0012\bar 3\bar 45678}} x_{A}^{\alpha _{\bar 3\bar 45678}} x_{B}^{-\alpha _{005678}}}{\Gamma \left(1+\alpha _1,\ldots,1+\alpha _8,1-\alpha _{005678},1+\alpha _{\bar 3\bar 45678},1-\alpha _{0012\bar 3\bar 45678},\alpha _{\bar 3\bar 4}-2\right)}\,,
\end{equation}

\begin{equation}
	P_1(x)=s_1 \left(\hat{x}_{12} x_{3A}+x_3 \left(\hat{x}_{2A}+\hat{x}_{49}\right)\right)+s_2 \hat{x}_{79} x_{3A}+s_5 \hat{x}_{89} x_{3A}+x_9 x_{3A}
	+\mu_1^2 x_9 \left(x_1 x_{3A}+\hat{x}_{3A}\right)+\mu_5^2\hat{x}_{29}  x_{3A},
\end{equation}
and
\begin{multline}
	G_1(x)=s_1 \left(\hat{x}_{12} x_{3A} x_{5678B}+\hat{x}_{369B}+\hat{x}_{235A}+\hat{x}_{3569}+x_{678B} \hat{x}_{23A}+\hat{x}_{349} x_{5678B}+\hat{x}_{569A}\right)
	\\
	+s_2 \hat{x}_{79AB}+s_3 \hat{x}_{89} \left(\hat{x}_{5A}+x_3 x_{5B}\right)+s_5 \hat{x}_{89AB}+s_4 \hat{x}_{679} x_{3A}+x_9 x_{3A} x_{5678B}\\
	+\mu_1^2 x_9 x_{5678B} \left(x_1 x_{3A}+\hat{x}_{3A}\right)+\mu_5^2 \hat{x}_{29} x_{3A} x_{5678B}.
\end{multline}
Then we find rescaling, which pulls out the dependence on $m_i$. It reads
\begin{equation}
	x_1\to \frac{x_1}{\mu_1^2},\ x_2\to \frac{x_2}{\mu_5^2},\ x_3\to \frac{x_3}{\mu_1^2},\ x_9\to \frac{x_9}{\mu_1^2 \mu_5^2},\ x_A\to \frac{x_A}{\mu_1^2}.
\end{equation}
which leads to the same form \eqref{eq:I63_80} of the contribution with the replacements $\mu_i\to 1$ and
\begin{equation}
	m^{2\alpha _{012\bar4\bar3\bar35678}}\to m_1^{2\alpha _{2\bar 3}} m_5^{2\alpha _{01\bar 3\bar 45678}}=v_3^{\alpha _{01\bar 3\bar 45678}} v_4^{\alpha _{2\bar 3}} s_1^{\alpha _{012\bar3\bar3\bar 45678}} s_2^{\alpha _{2\bar 3}} s_3^{\alpha _{\bar 0\bar 134\bar 5\bar 6\bar 7\bar 8}} s_4^{\alpha _{\bar 23}} s_5^{\alpha _{01\bar 3\bar 45678}}.
\end{equation}
Then we take successive limits $s_i\to 0$ at fixed $v_k$
\begin{equation}
	\left[\cI6{3,\text{dci}}\right]_{\#80}=\lim_{s_4\to0}\lim_{s_2\to0}\lim_{s_5\to0}\lim_{s_3\to0}\lim_{s_1\to0}\left[\cI6{3,\text{dci}}\right]_{\#80}
\end{equation}
to simplify the integral yet further. We obtain
\begin{equation}\label{eq:I63_80f}
	\left[\cI6{3,\text{dci}}\right]_{\#80}=v_3^{\alpha _{01\bar 3\bar 45678}} v_4^{\alpha _{2\bar 3}} \intop_{\mathbb{R}^{11}}dx_1\ldots dx_{9} dx_{A} dx_{B}
	f(x)\frac{P_2(x)}{G_2(x)^{3-\e}}
\end{equation}
where
\begin{align}
	G_2(x)
	&=x_{78} \hat{x}_{9AB}+\hat{x}_{23AB}+\hat{x}_{39}x_{5B}x_{2468A}+\hat{x}_{235A}+\hat{x}_{3679}+x_{29} \hat{x}_{13} x_{5B}+\hat{x}_{39}x_{5B},
	\\
	P_2(x)&=x_{78} \hat{x}_{39}.
\end{align}
Then we make rescaling $x_A \to x_1 x_A,\ x_8 \to x_7 x_8,\  x_5 \to x_B x_5$,
and take the integrals over $x_9$, $x_2$, $x_3$, $x_4$, $x_6$, $x_1$, $x_A$, $x_B$, $x_5$, $x_7$, $x_8$ (in this specific sequence). Finally we get
\begin{multline}
	\left[\cI6{3,\text{dci}}\right]_{\#80}=-v_3^{\alpha _{01\bar 3\bar 45678}} v_4^{\alpha _{2\bar 3}}\frac{\alpha _{03} \Gamma \left(1+\alpha _{0678},1+\alpha _{3\bar 6},1+\alpha _{012\bar 3\bar 45678}\right) }{\Gamma \left(1+\alpha _2,1+\alpha _3,1+\alpha _6,1+\alpha _7,1+\alpha _8\right)}
	\\
	\times\frac{\Gamma(\alpha _{\bar 23},-\alpha _{068},-\alpha _{078},\alpha _{\bar 368},\alpha _{0\bar 45678},-\alpha _{0\bar 35678},-\alpha _{01\bar 3\bar 45678})}{\Gamma(1-\alpha _{005678},1-\alpha_{0\bar3678},1+\alpha_{\bar 3\bar 45678},1+\alpha _{01\bar 45678},1-\alpha _{0012\bar 3\bar 45678})}\,.
\end{multline}
Proceeding in a similar way, we succeed in the calculation in terms of $\Gamma$-functions of all 54 regions which contribute to small-$m$ asymptotics of  $\cI63$. Finally, we added up contributions of different regions and took the limit $\alpha_i\to 0$ to obtain the slightly off-shell asymptotics of $\cI63$.

The same approach worked flawlessly also for integrals $\cI13,\ \cI23,\ \cI33$, $\cI73$, and $\cI83$.

\begin{figure}
	\centering
	\raisebox{-0.5\totalheight}{\includegraphics[height=4cm]{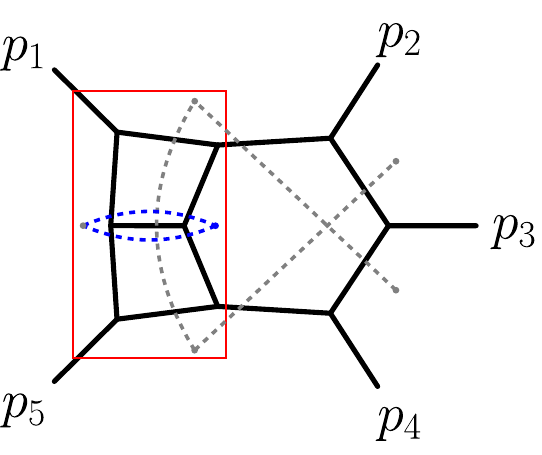}}
	\raisebox{-0.5\totalheight}{$\quad = \quad$}
	\raisebox{-0.5\totalheight}{\includegraphics[height=3.5cm]{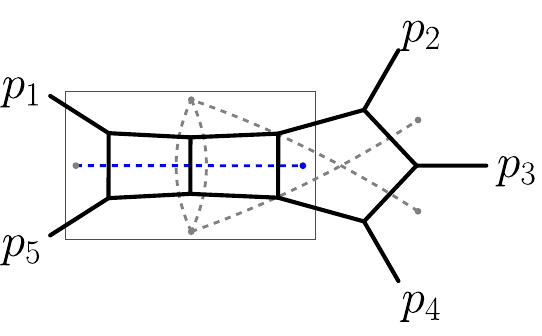}}
	\caption{Identity between integrals $\cI23$ and $\cI63$. The insertion of double box is shown in red frame.}
	\label{fig:I2=I6}
\end{figure}

The final results read
\begin{multline}\label{eq:I1-result}
	\cI13 =
	%I1
	\frac{1}{36} L_3^3
	\left(L_2+L_4\right)^3+\frac{1}{6}  L_3\, \left(L_2^2+3 L_3 L_2+2
	L_4 L_2+L_3^2+L_4^2+3 L_3 L_4\right)
	\left(L_2+L_4\right)\zeta _2 \\ +\frac{7}{2}\left(L_2^2+3 L_3 L_2+2
	L_4 L_2+L_3^2+L_4^2+3 L_3 L_4\right)\zeta _4+\frac{155 \zeta _6}{4}
	%I1/
	+O(v_i)\, ,
\end{multline}

\begin{multline}\label{eq:I3I6-result}
	\cI23=\cI63=
	%I2I6
	\frac{1}{12} L_1\, \left(L_3^2 L_2^3+3 L_3^2 L_4 L_2^2+3 L_3^2
	L_4^2 L_2+L_4^2 L_5^3+3 L_3 L_4^2 L_5^2+3 L_3^2 L_4^2 L_5\right)
	\\
	+\frac{1}{6}
	\big(L_1 L_2^3-L_3 L_2^3+6 L_1 L_3 L_2^2+3 L_1 L_4 L_2^2-3 L_3 L_4
	L_2^2+3 L_1 L_3^2 L_2  +3 L_1 L_4^2 L_2-3 L_3 L_4^2 L_2\\
	+12 L_1 L_3 L_4 L_2+L_1 L_5^3-L_4 L_5^3+3 L_3^2 L_4^2+3 L_1 L_3 L_4^2+3 L_1 L_3
	L_5^2 +6 L_1 L_4 L_5^2-3 L_3 L_4 L_5^2+3 L_1 L_3^2 L_4\\
	+3 L_1 L_3^2 L_5+3 L_1 L_4^2 L_5-3 L_3^2 L_4 L_5+12 L_1 L_3 L_4
	L_5\big)\zeta_2  -\frac{1}{6}  \big(L_2^3+3 L_4 L_2^2+3 L_4^2
	L_2+L_5^3+3 L_1 L_3^2\\
	+3 L_1 L_4^2-3 L_3 L_4^2+3 L_3 L_5^2-3 L_3^2 L_4+12 L_1 L_3 L_4+3 L_3^2 L_5\big)\zeta_3
	+\frac{1}{4} \big(-10 L_2^2+42
	L_1 L_2-10 L_3 L_2 \\-20 L_4 L_2+7 L_3^2  +7 L_4^2-10 L_5^2+35 L_1 L_3+35
	L_1 L_4+20 L_3 L_4 +42 L_1 L_5-20 L_3 L_5-10 L_4 L_5\big)\zeta_4 \\
	- (4 L_1 + L_2 - 5  L_3 - 5 L_4 + L_5)\zeta _2 \zeta _3 -4  L_1 \zeta_5
	+\frac{1}{8} \left(16 \zeta _3^2+77 \zeta _6\right)
	%I2I6/
	+O(v_i)
\end{multline}

\begin{multline}\label{eq:I3-result}
	\cI33 =
	%I3
	\frac{1}{12} \left(2 L_2 L_1^3+3 L_2 L_3 L_1^2+3 L_2 L_4
	L_1^2+6 L_2 L_3 L_4 L_1+L_3 L_4^3+3 L_2 L_3 L_4^2\right) L_5^2\\
	+\frac{1}{6}
	\big(2 L_2 L_1^3
	-2 L_5 L_1^3+3 L_5^2 L_1^2+3 L_2 L_3 L_1^2+3 L_2
	L_4 L_1^2+6 L_2 L_5 L_1^2-3 L_3 L_5 L_1^2 -3 L_4 L_5 L_1^2+3 L_2
	L_5^2 L_1\\+6 L_4 L_5^2 L_1
	+6 L_2 L_3 L_4 L_1+12 L_2 L_3 L_5 L_1-6
	L_3 L_4 L_5 L_1+L_3 L_4^3+3 L_2 L_3 L_4^2+3 L_4^2 L_5^2  +3 L_2 L_3
	L_5^2\\+3 L_2 L_4 L_5^2 +3 L_3 L_4 L_5^2-L_4^3 L_5-3 L_2 L_4^2 L_5+12
	L_2 L_3 L_4 L_5\big) \zeta_2 -\frac{1}{6} \big(2 L_1^3+3 L_3
	L_1^2+3 L_4 L_1^2\\+6 L_3 L_4 L_1-L_4^3
	-3 L_2 L_4^2-3 L_2 L_5^2+6 L_4 L_5^2-6 L_2 L_4 L_5-6 L_3 L_4 L_5\big) \zeta_3
	+\frac{1}{4} \big(10 L_4^2+20 L_1 L_4\\-11 L_2 L_4+L_3
	L_4+20 L_5 L_4+10 L_5^2+10 L_1 L_2-20 L_1 L_3+42 L_2 L_3+30 L_1
	L_5  +6 L_2 L_5+4 L_3 L_5\big)\zeta_4  \\+2 (L_2+  L_3)\zeta _5
	-\big(3  L_5 + L_1-3 L_2-L_3+4 L_4\big)\zeta _2 \zeta _3   -3 \zeta _3^2+28 \zeta _6
	%I3/
	+O(v_i)
\end{multline}

\begin{multline}\label{eq:I7-result}
	\cI73 =
	%I7
	\frac{1}{12} L_1 L_2\, \left(2 L_1^2+3 L_3 L_1+3 L_4 L_1+6 L_3
	L_4\right) L_5\, \left(L_2+L_5\right)
	+\frac{1}{2}  \big(L_2^2 L_1^2+L_5^2 L_1^2
	\\+4 L_2 L_5 L_1^2+L_2 L_5^2 L_1+2 L_4 L_5^2 L_1+2
	L_2^2 L_3 L_1+L_2^2 L_5 L_1  +4 L_2 L_3 L_5 L_1+4 L_2 L_4 L_5 L_1\\+L_3
	L_4 L_5^2+L_2^2 L_3 L_4+4 L_2 L_3 L_4 L_5\big)\zeta_2 -\frac{1}{6}  \big(4 L_1^3+6 L_3 L_1^2+6 L_4 L_1^2
	+12 L_3 L_4 L_1-6 L_2
	L_5^2\\
	-3 L_3 L_5^2+3 L_4 L_5^2+3 L_2^2 L_3-3 L_2^2 L_4 +6 L_2 L_3
	L_4-6 L_2^2 L_5+6 L_3 L_4 L_5\big)\zeta_3+\frac{1}{4}
	\big(3 L_2^2+40 L_1 L_2\\
	+31 L_3 L_2-L_4 L_2+12 L_5 L_2 +3 L_5^2+40
	L_1 L_5-L_3 L_5+31 L_4 L_5\big)\zeta_4
	-\big(2  L_1+3 L_2+3 L_5\big) \zeta _2 \zeta _3 \\  +2 (L_2+L_5)\zeta _5
	-4 \zeta _3^2+28\zeta _6
	%I7/
	+O(v_i)
\end{multline}

\begin{multline}
	\cI83 =
	%I8
	-\frac{1}{3}  \left(L_1+L_4\right)^3\zeta _3+3\left(L_1+L_4\right)^2 \zeta _4 -2  \left(L_1+L_4\right)\zeta _2 \zeta _3
	-10 \left(L_1+L_4\right) \zeta _5
	-4 \zeta _3^2+\frac{41 \zeta _6}{2}
	\\
	%I8/
	+O(v_i)
\end{multline}
Several remarks are in order. First, for all but one of the above integrals, the contributions of all regions were expressed in terms of $\Gamma$ functions. The only exception appeared to be the simplest integral $\cI83$, where the contributions of the 4 nonzero regions were expressed in terms of hypergeometric functions. Next, the integrals $\cI23$ and $\cI63$ are equal to each other due to the magic identity for double box, see Fig. \ref{fig:I2=I6}. They have been considered independently within the described approach, and the identity of the two results serves as a good cross-check.
Another surprise appeared in the calculation of $\cI43$ and  $\cI{4'}3$ integrals. It turns out that for these two integrals, some of the contributions are at least hypergeometric. But given the exact identity $\cI43=\cI{4'}3=\cI13$, we were not motivated enough to elaborate on these two integrals.

\section{Reducing to locally finite integrals}

\begin{figure}
	\centering
	\includegraphics[width=0.27\linewidth]{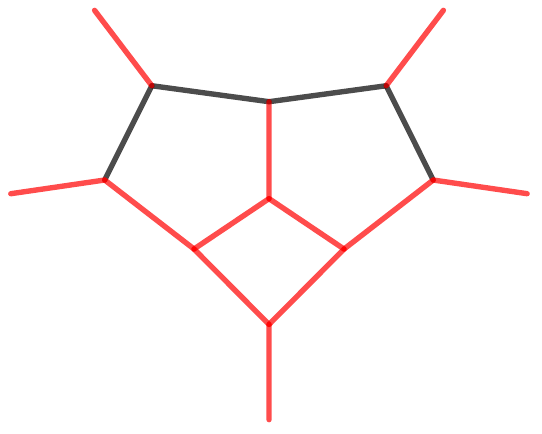}
	\includegraphics[width=0.27\linewidth]{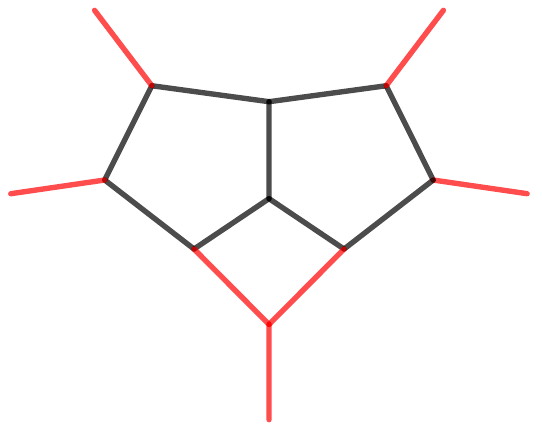}
	\includegraphics[width=0.27\linewidth]{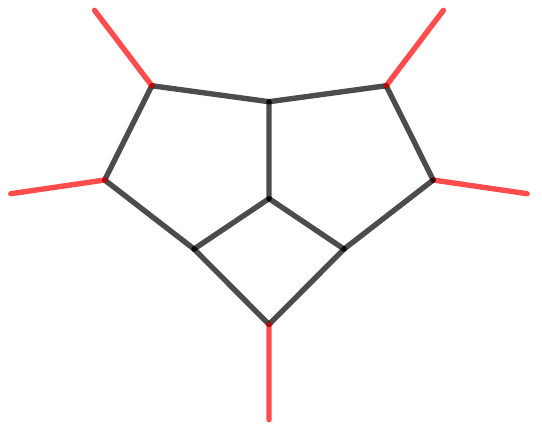}
	\caption{Problematic regions \#\#99,162,163 of the integral $\cI53$.}
	\label{fig:I5hardregions}
\end{figure}

Although the approach in the previous section worked perfectly for the integrals $\cI23$, $\cI33$, $\cI63$, $\cI73$, it was not sufficient to calculate $\cI53$. Nevertheless, we were able to pull the kinematic factor out of the integral by the method described in the previous section. It is already at this stage that we are guaranteed that $\cI53$, as a function of cross ratios, is expressed in terms of the logarithms only. However, the parametric integrals which remained to be taken prove to be not always expressible in terms of $\Gamma$-functions. More precisely, out of 163 regions, we find 81 regions with zero contribution, 67 regions with contribution expressible via $\Gamma$-functions, 12 regions with contribution expressible via $_{p}F_{q}(\ldots|1)$, and 3 regions (\#\#99,162,163) for which we could not find a decent representation. These three regions are depicted in Fig. \ref{fig:I5hardregions}.

We were able to derive 6-, 5-, and 5-fold integral representations over Feynman parameters for region \#99, \#162, and \#163, respectively.

The general form of the obtained representations can be presented as

\begin{equation}\label{eq:parametric}
	\intop_{\mathbb{R}^N}dxx^{\nu-1}G(x)^{\mu} P(x)\stackrel{def}{=}
	\intop_{\mathbb{R}^N}dx_1\ldots dx_N\  x_1^{\nu_1-1}\ldots x_N^{\nu_N-1} G_1(x)^{\mu_1}\ldots G_M(x)^{\mu_M} P(x),
\end{equation}
where $\nu_k$ and $\mu_l$ are some linear functions of regularization parameters, $G_l$ are some polynomials of $x_1\ldots x_n$, and $P(x)$ is a polynomial of $x_1\ldots x_n$ and  regularization parameters.

From the practical point of view, we would like to take all regularization parameters as $\a_k=p_k\e$, where $p_k$ are rather generic integer coefficients. Then we would like to expand the integrand in $\e$ and use \HI to obtain the expansion of the integral in terms of polylogarithmic constants. However, this approach works only for the locally finite integrals, for which the $\e$-expansion commutes with the integration. Unfortunately, the integral representations for our three problematic regions are not \emph{a priori} locally finite. Thus, we need to somehow transform the initial representation to a locally finite form.

There are several constructive approaches to achieve this goal existing in the literature. They rely on an understanding of the tropical geometry of parametric integrands.
It is important that for our present application the polynomials $G_l$ always have \emph{positive} coefficients. It means that the divergences may emerge only from the boundaries of the integration regions; none of them can arise from the bulk. This allows one to systematically analyze divergences by considering the Newton polytope of the integrand, $\operatorname{Newt}(G)=\operatorname{Newt}(\prod_lG_l)=\sum_l \operatorname{Newt}(G_l)$ (where $\sum_l$ denotes Minkowski sum). We remind that the Newton polytope of the polynomial $g(x)=\sum_r c_r x^{n^{(r)}}=\sum_r c_r x_1^{n_1^{(r)}}\ldots x_N^{n_N^{(r)}}$ is defined as the convex hull of the exponent vectors of its monomials,
\begin{equation}
	\operatorname{Newt}(g) = \operatorname{conv}(n^{(1)},\ldots, n^{(R)}).
\end{equation}
The outer normals of the faces of $\operatorname{Newt}(G)$ determine the rays along each of which the scaling behavior of the integrand needs to be considered for the analysis of divergences. Namely, let $n$ be an outer normal vector for some face of  $\operatorname{Newt}(G)$. Then we rescale $x\to \lambda^n x$ and consider the limit $\lambda\to\infty$. In this limit the integrand scales as $\lambda^{\sigma}$, where the exponent is a linear function of regularization parameters, $\sigma=\sigma(\alpha)$. If for all rays we have $\sigma(\alpha=0)<0$, the integral is locally finite. Otherwise the integral contains divergences. These need to be eliminated to make the integral locally finite. There are two ``tropical'' ways to accomplish this.

The first route uses integration by parts along divergent rays. This is known as the Nilsson-Passare analytic continuation \cite{Nilsson}.  It is similar
in spirit to the integration by parts in divergent momentum integrals of 't Hooft and Veltman in their seminal application of dimensional regularization
to Feynman integrals \cite{tHooft:1972tcz}. Notice, however, that this method results in a significant proliferation of integrals to be calculated, as one
typically gets twice or more integrands at each integration step. For a high-dimensional ambient space for Newton polytopes and high-dimensional fans of divergent rays, calculations become impractical.

Another route is to `subtract' divergent directions from integrands. This can be done systematically, as was demonstrated by Salvatori
\cite{Salvatori:2024nva} (see also \cite{Belitsky:2025sin}). The advantage of this method over the Nilsson-Passare procedure is that one can
explicitly extract poles along divergent rays via barycentric change of Feynman parameters \cite{Kaneko:2009qx}, on the one hand, and reduce
the dimension of accompanying remainder integrals, on the other. However, if the divergent fan is not simplicial, linearly dependent rays need
to be integrated out via the above Nilsson-Passare procedure. So, again, as in the first method, for divergent fans of high dimension, the
inclusion-exclusion procedure and the need to deal with simplicial divergent fans only yield a profusion of integrals to be analyzed and fed into the \HI.

These two ways of achieving local finiteness were successfully applied in \cite{Belitsky:2025sin}. However, because of the complications just described, we find it more practical to use, in the present more complicated situation, the third approach based on the IBP reduction in parametric representation. The basic idea is very simple. First, we generate a sufficiently large set of integrals of the form
\begin{equation}
	\intop_{\mathbb{R}^N}dxx^{\nu+n-1}G(x)^{\mu+m},
\end{equation}
where $n$ and $m$ denote integer shifts of indices. In practice we use the set of integrals with shifts $(n_1,\ldots,n_N,m_1,\ldots m_M)\in \{-1,0,1\}^{N+M}$
Then we pick out of this set the locally finite integrals using the criterion described above. Finally, we use IBP reduction in parametric space to reduce the original integral to the master integrals judiciously chosen from the set of locally finite integrals.

Let us mention that, for Feynman integrals considered as parametric integrals,
the authors of Ref.~\cite{vonManteuffel:2014qoa} used
a closely related notion of {\em quasi-finite} Feynman integrals
which meant that the only source of singularities of such integrals
in $\varepsilon$ can be an overall gamma function.
Obviously, any quasi-finite integral is also locally finite.
The goal of finding a basis consisting of quasi-finite master integrals
was similar to ours: to obtain the possibility to perform an
expansion in $\varepsilon$ under parametric integral sign and then to apply
\HI. An important point of the strategy of  Ref.~\cite{vonManteuffel:2014qoa} was to
look for quasi-finite integrals in shifted dimensions \footnote{
	The method of Ref.~\cite{vonManteuffel:2014qoa} was successfully applied, for example, to the evaluation of four-loop form-factor integrals ---	see, e.g., \cite{Lee:2023dtc}.}.

\subsection{IBP reduction in parametric representation}

Let us explain how we devise the IBP reduction in parametric representation.
We consider the infinite family of integrals
\begin{equation}
	j(n,m)=j(n_1,\ldots,n_N,m_1,\ldots, m_M)=	\intop_{\mathbb{R}^N}dxx^{\nu+n-1}G(x)^{\mu+m},
\end{equation}
which depend on integer indices $n_1,\ldots n_N,m_1,\ldots,m_M$.

We have two kinds of relations between integrals of this family. First, we have integration by parts relations which follow from the identity
\begin{equation}
	0=\intop_{\mathbb{R}^N}dx \partial_{k} x^{\nu+n-1}G(x)^{\mu+m},
\end{equation}
where $\partial_{k}=\frac{\partial}{\partial x_{k}}$.
Expressing the result of explicit differentiation in terms of the integrals of the same family, we obtain recurrence relations between the integrals.

Another kind of relations follows from the algebraic identity
\begin{equation}\label{eq:mshifts}
	\intop_{\mathbb{R}^N}dx x^{\nu+n-1}G(x)^{\mu+m-\delta^{(l)}}G_l(x) =\intop_{\mathbb{R}^N}dx x^{\nu+n-1}G(x)^{\mu+m},
\end{equation}
Here $\delta_{l'}^{(l)}=\delta_{l'l}$, so that  $G(x)^{\mu+m-\delta^{(l)}}=G(x)^{\mu+m}/G_l(x)$. Expressing $G_l(x)$ as a sum of monomials, we obtain in the left-hand side a linear combination of integrals with shifted indices $n$.

This approach, however, obliges us to consider the family depending on $N+M$ indices, a number which might essentially exceed the number of integration variables.

Meanwhile, there is a simple way to obtain relations between integrals belonging to a slice of constant $m$. This approach is a simple generalization of the one described in Ref. \cite{Lee:2014tja}. The idea is to construct a system of polynomial first-order differential operators of the form $Q(x) \cdot \partial \equiv \sum_kQ_k(x) \partial_{k}$ with the following property:
\begin{equation}\label{eq:ibp_operator}
	Q(x) \cdot \partial\, G_l \text{ is divisible by } G_l,\quad (l=1,\ldots,M)
\end{equation}
Here, $Q_k(x)$ are some polynomials. We note en passant that such differential operators have a nice geometric meaning as tangent vector fields on the variety defined by polynomial equations $G_1(x)=\ldots=G_M(x)=0$. Obviously, they form a (left) module over the ring of multivariate polynomials of $x$. It is worth noting that the basis of this module can be found using routine commutative algebra algorithms implemented, e.g., in \texttt{Singular} CAS \cite{DGPS}. Namely, we first find for each $l$ the syzygy module of $\langle\partial_1G_l,\ldots, \partial_NG_l, G \rangle$. Omitting the last component of each syzygy, we obtain the module of tangent vector fields on the variety $G_l=0$. Then, calculating the intersection of these modules for all $l$, we find the required module.

Now, each differential operator $	Q(x)\cdot \partial$, satisfying \eqref{eq:ibp_operator}, generates the IBP identity
\begin{equation}
	0=\intop_{\mathbb{R}^N}dx \,\partial \cdot \left[Q(x) x^{\nu+n-1}G(x)^{\mu+m}\right]
\end{equation}
which does not shift $m$ indices. In this approach the identities \eqref{eq:mshifts} play the role of dimension shifting relations in the conventional approach.

\subsection{Integral $\cI53$}
Applying the described approach to the obtained parametric representations for the three problematic regions of $\cI53$, we have successfully expressed their contributions via locally finite integrals treatable by \HI. Unfortunately, the computational complexity did not allow us to calculate a few of the highest coefficients in $\e$-expansion. We have calculated the expansions up to $\e^{-2}$, $\e^{0}$, $\e^{-1}$ for the regions \#99, \#162, and \#163, respectively.
In order to further restrict the uncertainty, we used the condition of finiteness of $\cI53$. This condition, in particular, requires that the coefficients in front of each monomial in $L_i$ are finite. Finally, we were able to find the result for $\cI53$ up to a constant independent of $L_i$. As this result has to be uniform transcendental, and expressible in terms of zeta functions and logarithms only, this constant has the form $c_1\zeta_6+c_2\zeta_3^3$, where $c_1$ and $c_2$ are rational numbers. Our final result for $\cI53$ reads

\begin{multline}
	\cI53=
	%I5
	\frac{1}{12} \big(L_5^2 L_1^3+L_2 L_5 L_1^3+3 L_4 L_5^2 L_1^2+3 L_2 L_4 L_5 L_1^2+6 L_3 L_4 L_5^2 L_1+3 L_2 L_3^2 L_4 L_1+L_2 L_3^3 L_4\\
	+6 L_2 L_3 L_4 L_5 L_1\big) L_2
	+\frac{1}{6} \big(-L_2 L_3^3+L_4 L_3^3-3 L_1 L_2 L_3^2+3 L_1 L_4 L_3^2+6 L_2 L_4 L_3^2+6 L_1 L_2^2 L_3-3 L_1 L_5^2 L_3\\
	-3 L_2 L_5^2 L_3+3 L_4 L_5^2 L_3+3 L_2^2 L_4 L_3+6 L_1 L_2 L_4 L_3-3 L_2^2 L_5 L_3+12 L_2 L_4 L_5 L_3+3 L_1^2 L_2^2+3 L_1 L_2 L_5^2\\
	+6 L_1 L_4 L_5^2+3 L_1 L_2^2 L_4+3 L_1 L_2^2 L_5+6 L_1^2 L_2 L_5+12 L_1 L_2 L_4 L_5\big)\zeta _2
	+ \frac{1}{6} \big(6 L_5^2 L_3-L_3^3-3 L_1 L_3^2+3 L_2^2 L_3\\
	+6 L_1 L_2 L_3-6 L_1 L_4 L_3-12 L_2 L_4 L_3+6 L_1 L_5 L_3
	-6 L_4 L_5 L_3-6 L_1 L_2^2+3 L_1 L_5^2+3 L_2 L_5^2-3 L_4 L_5^2\\
	+6 L_2^2 L_4+6 L_1 L_2 L_4+3 L_2^2 L_5\big)\zeta _3 +\frac{1}{4} \big(3 L_2^2+41 L_1 L_2+20 L_3 L_2+L_4 L_2+28 L_5 L_2-10 L_3^2-4 L_5^2\\
	+12 L_1 L_3+9 L_1 L_4+9 L_3 L_4+9 L_1 L_5-22 L_3 L_5+31 L_4 L_5\big)\zeta _4 + \big(7 L_3-4 L_1-7 L_2+2 L_4-4 L_5\big)\zeta _2 \zeta _3\\
	-2  \big(3 L_1+L_2+2 L_3-2 L_5\big)\zeta _5+c_1 \zeta _6+ c_2 \zeta _3^2
	%I5/
	+O(v_i)
\end{multline}

\section{Conclusion}

In the present paper, we have calculated the three-loop five-point integrals relevant for the amplitude in the planar limit of $\mathcal{N}=4$ SYM theory. Apart from the most complicated integral $\cI53$, we have been able to obtain complete results. The result for the integral $\cI53$ lacks the constant term, which should conjecturally be a linear combination of $\zeta_6$ and $\zeta_3^2$. The obtained results are provided in ancillary files in computer-readable form.

The simplicity of our results — expressed solely through logarithms of cross-ratios — is a consequence of the retained dual conformal symmetry. This is in contrast with the only other known computations of three-loop five-point integrals, \cite{Liu2025,Chicherin2025}, performed in dimensional regularization for strictly on-shell integrals. Those results are technically much more involved, but a direct comparison with our slightly off-shell results is not possible due to the different kinematic regimes and regularization schemes.

In light of the results presented in this work, the computation of four-loop five-point integrals may no longer appear to be an entirely hopeless enterprise, although it would definitely require a more systematic integration of the concepts discussed above, along with comprehensive automation of all computational stages. Another possible direction in which our manifestly DCI-invariant approach to MofR can definitely be useful already in its current form is the two-loop evaluation of six-point DCI integrals contributing to the six-point amplitude in $\mathcal{N}=4$ SYM on the special Coulomb branch \cite{Belitsky:2025vfc}.
\begin{acknowledgments}
	We are grateful to Erik Panzer for his valuable advice regarding the application of {\tt HyperInt}. The work of V.S. was supported
	by the Moscow Center for Fundamental and Applied Mathematics of Lomonosov Moscow State University under Agreement No.\ 075-15-2025-345. L.B.\ was supported by the Foundation for the Advancement of Theoretical Physics and Mathematics ``BASIS''.
\end{acknowledgments}

	\bibliographystyle{apsrev4-2}
	\bibliography{penta3loopPRD}
	
\end{document}

%% file: commands.tex
\newcommand{\e}{\epsilon}
\renewcommand{\a}{\alpha}

\newcommand{\cI}[2]{\mathcal{I}_{#1}^{(#2)}}
\newcommand{\HI}{\texttt{HyperInt}\xspace}